\documentclass{PoS}
\usepackage{amsmath,amssymb,epsfig,cite}

\newcommand{\myScript}[1]{\mathcal{#1}}

\newcommand{\Equation}[1]{Eq.~(\ref{#1})}

\newcommand{\ie}{{\it i.e.}}

\newcommand{\Ord}{\myScript{O}}

\newcommand{\imag}{\mathrm{i}}

\newcommand{\dDell}{[d\ell]}

\newcommand{\leftA}[1]{\langle#1|}

\newcommand{\Srght}[1]{|#1]}
\newcommand{\lop}[2]{#1\!\cdot\!#2}

\newcommand{\Den}{\mathcal{D}}

\newcommand{\BoxC}{C}
\newcommand{\TriC}{C}
\newcommand{\BubC}{C}
\newcommand{\GenPol}{\mathcal{C}}

\newcommand{\tBoxPol}{\tilde{\GenPol}}
\newcommand{\tTriPol}{\tilde{\GenPol}}
\newcommand{\tBubPol}{\tilde{\GenPol}}
\newcommand{\graph}[3]{\raisebox{-#3ex}{\epsfig{file=#1.pdf,width=#2ex}}}

\title{%
\vspace{-10.6ex}{\normalsize\normalfont\hspace{\fill}IFJPAN-IV-2018-13}\\
\vspace{8ex}%
One-loop amplitudes for $\mathbf{k}_\mathbf{T}$-dependent factorization%
}

\ShortTitle{One-loop amplitudes for $k_T$-dependent factorization}

\author{\speaker{Andreas van Hameren}\,\thanks{This work was supported by grant of National Science Center, Poland, No. 2015/17/B/ST2/01838.}\\
        Institute of Nuclear Physics Polish Academy of Sciences, Krak\'ow\\
        E-mail: \email{hameren@ifj.edu.pl}}

\abstract{I report on progress in the calculation of one-loop amplitudes with a space-like gluon. Amplitudes with space-like partons are relevant for factorization procedures that require initial-state partons with non-vanishing transverse momentum components.}

\FullConference{Loops and Legs in Quantum Field Theory (LL2018)\\
		29 April 2018 - 04 May 2018\\
		St.\ Goar, Germany}

\begin{document}

\section{Introduction}
Factorization prescriptions in which one or both initial-state partons have an explicit dependence on the transverse momentum ($k_T$) require in general partonic hard matrix elements with space-like initial-state partons.
For example $k_T$-factorization~\cite{Gribov:1984tu} and high-energy factorization~\cite{Catani:1990eg,Collins:1991ty} are of this type, and so is the more recent example of improved transverse momentum dependent factorization~\cite{Kotko:2015ura}.
Preservation of gauge invariance demands some special care regarding the definition and calculation of the partonic matrix elements.
Several approaches leading to equivalent results at tree-level exist~\cite{Lipatov:1995pn,Lipatov:2000se,Antonov:2004hh,vanHameren:2012if,vanHameren:2013csa,Kotko:2014aba,Karpishkov:2017kph}, and some even allow for efficient ``on-shell recursion'', leading expressions that are rather compact and remarkably similar to those for fully on-shell amplitudes~\cite{vanHameren:2014iua,vanHameren:2015bba}.
The completely automated parton-level Monte Carlo program \mbox{\sc Ka\hspace{-0.2ex}Tie}~\cite{vanHameren:2016kkz} can perform tree-level calculations for arbitrary processes within the Standard Model.
It can be used in combination with TMDlib~\cite{Hautmann:2014kza} to produce fully exclusive parton-level event files that can be used in combination with parton-shower programs~\cite{Bury:2017jxo}.

This type of calculations must be advanced to next-to-leading order (NLO) in the perturbative expansion, both in order to reach higher precision and to confirm the reliability of the factorization procedure.
The possibility of a systematic treatment of the divergences that accompany higher-order calculations within quantum chromodynamics must emerge for the factorization procedure to be physically relevant.
The so-called parton reggeization approach has allowed for obtaining some results already~\cite{Hentschinski:2011tz,Chachamis:2012cc,Chachamis:2013hma,Hentschinski:2014lma,Nefedov:2016clr,Nefedov:2017qzc}.
Other recent NLO calculations with explicit $k_T$ can be found in~\cite{Boussarie:2016ogo,Beuf:2017bpd}, and more specifically related to the color-glass-condensate formalism in~\cite{Iancu:2016vyg,Ducloue:2017mpb}, and applying light-cone perturbation theory in~\cite{Hanninen:2017ddy,Lappi:2016oup}.

This write-up reports on the advancement of the method of~\cite{vanHameren:2012if} to define and calculate tree-level amplitudes to one-loop amplitudes.
It achieves the required space-like nature of the initial-state gluons by interpreting them as an auxiliary on-shell quark-antiquark pair with particularly chosen light-like momenta.
This way, it manifestly leads to gauge invariant amplitudes.
Considering only the kinematics, there is a freedom of choice of the momenta of the auxiliary quarks, expressed by a parameter $\Lambda$.
Dividing the amplitude by $\Lambda$ for each auxiliary pair and taking $\Lambda$ to infinity leads to an amplitude that can be calculated directly with the application of eikonal Feynman rules for the auxiliary quark lines.
The appropriateness of this procedure is highlighted by the fact that it removes a color-degree of freedom for each auxiliary pair, giving them the color content of single gluons, as required.
This procedure is rather simple for tree-level amplitudes because they are just rational functions in the (Weyl spinors of the) momenta of the external partons.
This is not the case anymore for one-loop amplitudes, and some care needs to be taken.
In particular, it turns out that the parameter $\Lambda$ acts as a regulator for divergencies that are well-known to occur in amplitudes with so-called linear denominators stemming from the eikonal Feynman rules.

\section{Amplitudes with a space-like gluon}
In the factorized picture the hadron-scattering the cross section is calculated as a convolution of a parton-level hard scattering matrix element with parton distribution functions.
The $k_T$-dependent space-like initial-state gluon is required to have momentum 
%
\begin{equation}
k^\mu = xp^\mu + k_T^\mu
\quad,
\end{equation}
%
where $p^\mu$ is the light-like momentum of one of the colliding hadrons, and $\lop{p}{k_T}=0$.
The amplitude is defined with the help of an auxiliary quark-antiquark pair via the association
%
\begin{equation}
\graph{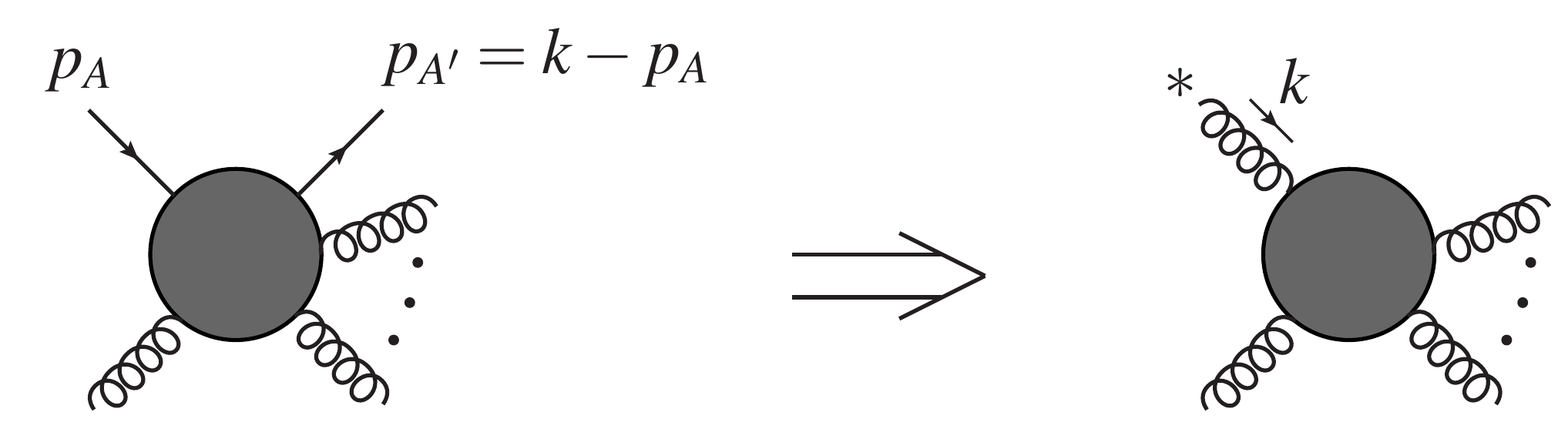}{50}{5.5}
\end{equation}
%
where both $p_{A}^\mu$ and $p_{A'}^\mu$ are light-like.
More explicitly, an arbitrary light-like momentum $q^\mu$ can be chosen such that $\lop{q}{k_T}=0$ while $\lop{q}{p}\neq0$ and one can define
%
\begin{equation}
p_{A}^\mu = \Lambda p^\mu + \alpha q^\mu + \beta k_{T}^\mu
\quad\textrm{with}\quad
\alpha = \frac{-\beta^2k_{T}^2}{\Lambda(p+q)^2}
\quad\textrm{and}\quad
\beta = \frac{1}{1+\sqrt{1-x/\Lambda}}
\quad.
\label{Eq:pA}
\end{equation}
%
This way, for any value of $\Lambda$, both $p_A^\mu$ and $p_{A'}^\mu$ are guaranteed to be light-like and sum up to $p_{A}^\mu+p_{A'}^\mu=k^\mu$.
The desired amplitude is obtained by taking $\Lambda$ to infinity:
%
\begin{equation}
\frac{|k_T|}{\Lambda}\,\mathcal{A}\big(\emptyset\to\bar{\mathrm{q}}(p_{A}(\Lambda))\,\mathrm{q}(p_{A'}(\Lambda))+X\big)
\;\;\overset{\Lambda\to\infty}{\longrightarrow}\;\;
\mathcal{A}\big(\emptyset\to\mathrm{g}^*(xp+k_T)+X\big)
\quad.
\label{limit}
\end{equation}
%
As stated before, the limit it trivial for tree-level amplitudes, and the limiting amplitude can be obtained directly with the help of eikonal Feynman rules.
For one-loop amplitudes, the limit is not that straightforward.
At the level of the one-loop integrand, taking the limit $\Lambda\to\infty$ will result in the appearance of linear denominators, which lead to divergent integrals that need to be regularized.
A regularization is automatically provided by not taking the limit on the integrand, but rather after integration.
Because of the known existence of the divergencies, this then must necessarily lead to divergencies parametrized by $\Lambda$.

Loop amplitudes are in general calculated within dimensional regularization in order to deal with the various types of divergencies that occur.
The amplitude becomes a function of the parameter $\varepsilon$, deforming the integration dimension from $4$ to $4-2\varepsilon$, and is written as a Laurent expansion in terms of this parameter.
The calculation of a cross section can only be considered physically relevant if all possible poles in $\varepsilon$, coming from real contributions, virtual contributions, and renormalization contributions, eventually  cancel.
Extracting the Laurent expansion from the amplitude means taking the limit $\varepsilon\to0$.
The regularization of the linear denominators by means of $\Lambda$ gives the opportunity to take the limit $\Lambda\to\infty$ after taking the limit of $\varepsilon\to0$, \ie\ taking expansion in $\varepsilon$ and expand it in (logarithms of) $\Lambda$.
This has the advantage that existing expressions for on-shell amplitudes with quark-antiquark pairs can be used directly to obtain expressions for amplitudes with space-like gluons.
This approach would become cumbersome for amplitudes with many external partons, which one would rather like to obtain using an expansion in terms of master integrals with numerically calculable coefficients, to avoid large algebraic expressions.
For this approach, expressions for one-loop master integrals can simply be expanded in (logarithms of) $\Lambda$, and no new loop integrals need to be calculated.
The question is then if this expansion can be organized such that the coefficients can be calculated at the limit of $\Lambda\to\infty$.
Preceding this question is, however, the question whether one-loop amplitudes can be defined with \Equation{limit}, that is, whether the divergencies in $\Lambda$ are at most logarithmic and not power-like.

\section{Integrand-level reduction methods}
These questions have been answered in detail in~\cite{vanHameren:2017hxx}, and the main findings are reported below.
For the purpose of explaining the solved issues the well-known integrand-level methods for one-loop amplitudes must be briefly addressed.

In order to avoid complications that are irrelevant for the particular problem being studied, the analysis is restricted to {\em primitive amplitudes\/}~\cite{Bern:1994fz}.
They are planar and can conveniently be written as the integral of a single numerator with a complete set of all possible loop-momentum dependent propagator denominators
%
\begin{equation}
\Den_i(\ell) = (\ell+K_i)^2+\imag\eta
~,
\end{equation}
%
occuring in the the contributing one-loop graphs.
The denominator momentum $K_i^\mu$ is a sum of a subset of external momenta, and $\eta$ is small and positive in order to enforce the Feynman prescription.
The decomposition of such an amplitude in terms of {\em master integrals\/} looks like
%
\begin{multline}
\int \dDell\,\frac{\mathcal{N}(\ell)}{\prod_i\Den_i(\ell)}
= \sum_{i,j,k,l} \BoxC_{ijkl}\,\mathrm{Box}_{ijkl}
             + \sum_{i,j,k} \TriC_{ijk}\,\mathrm{Tri}_{ijk}
             + \sum_{i,j} \BubC_{ij}\,\mathrm{Bub}_{ij}
             + \mathcal{R} + \mathcal{O}(\varepsilon)
~,
\label{oneloopdecom}
\end{multline}
%
where the master integrals are defined as
%
\begin{align}
\mathrm{Box}_{ijkl}= \int\frac{\dDell}{\Den_i(\ell)\Den_j(\ell)\Den_k(\ell)\Den_l(\ell)}
\;,\;\;
\mathrm{Tri}_{ijk} = \int\frac{\dDell}{\Den_i(\ell)\Den_j(\ell)\Den_k(\ell)}
\;,\;\;
\mathrm{Bub}_{ij} = \int\frac{\dDell}{\Den_i(\ell)\Den_j(\ell)}
~.
\end{align}
%
We only consider massless denominators here, so tadpole integrals do not contribute.
The sums are over all non-equal values of the indices, and the ``normalization'' of the dimensionally regulated loop volume element is given by
%
\begin{equation}
\dDell=
\frac%
{\Gamma(2-\varepsilon)\mu^{2\varepsilon}}
{\Gamma^2(1-\varepsilon)\Gamma(1+\varepsilon)\imag\pi^{2-\varepsilon}}
\,d^{4-2\varepsilon}\ell
~.
\end{equation}
%
The master integrals contain (poly)-logarithms of rational functions of external momenta, and $\mathcal{R}$ represents the rational terms.

Following the method of~\cite{Ossola:2006us,Ellis:2007br}, the coefficients in the expansion can be found from the integrand expansion
%
\begin{equation}
\frac{\mathcal{N}(\ell)}{\prod_i\Den_i(\ell)}
= \sum_{i,j,k,l}\frac{C_{ijkl}+\tBoxPol_{ijkl}(\ell)}
                     {\Den_{i}(\ell)\Den_{j}(\ell)\Den_{k}(\ell)\Den_{l}(\ell)}
+ \sum_{i,j,k}\frac{C_{ijk}+\tTriPol_{ijk}(\ell)}
                   {\Den_{i}(\ell)\Den_{j}(\ell)\Den_{k}(\ell)}
+ \sum_{i,j}\frac{C_{ij}+\tBubPol_{ij}(\ell)}
                 {\Den_{i}(\ell)\Den_{j}(\ell)}
+ \sum_{i}\frac{\tBubPol_{i}(\ell)}
                 {\Den_{i}(\ell)}
~,
\label{integranddecom}
\end{equation}
%
which, besides the desired ones, involves more coefficients that are implicit in the polynomials denoted with a tilde.
The relation holds for any value of the integration momentum $\ell^\mu$, and thus implies an infinite set of equations from which all coefficients can be solved.
Rather than choosing arbitrary values of $\ell^\mu$ to evaluate the relation, it is extremely useful to consider values for which denominators vanish.
Choosing $\ell^\mu$ such that $\Den_{i}(\ell)=\Den_{j}(\ell)=\Den_{k}(\ell)=\Den_{l}(\ell)=0$, which is possible in $4$ dimensions, reduces the relation to one involving only the {\em box coefficient\/} $C_{ijkl}$ and {\em spurious box coefficients\/} implicit in $\tBoxPol_{ijkl}(\ell)$.
Having solved for all box coefficients, the box terms in \Equation{integranddecom} can be moved to the left-hand-side (LHS), and the same procedure can be applied to the triangle terms, etc.

This approach leads to the, also extremely useful, unitarity interpretation.
Choosing $\ell^\mu$ such that denominators vanish amounts to {\em cutting\/} internal lines in the graphs representing the LHS of \Equation{integranddecom}, that is putting them on-shell.
Evaluating the LHS at $\ell^\mu$ such that denominators $i,j,k,l$ vanish, after multiplying with those denominators, means considering the simultaneous residue with respect to these denominators.
This residue consists of the graphs that actually contain all four denominators, and can be depicted by
%
\begin{equation}
\graph{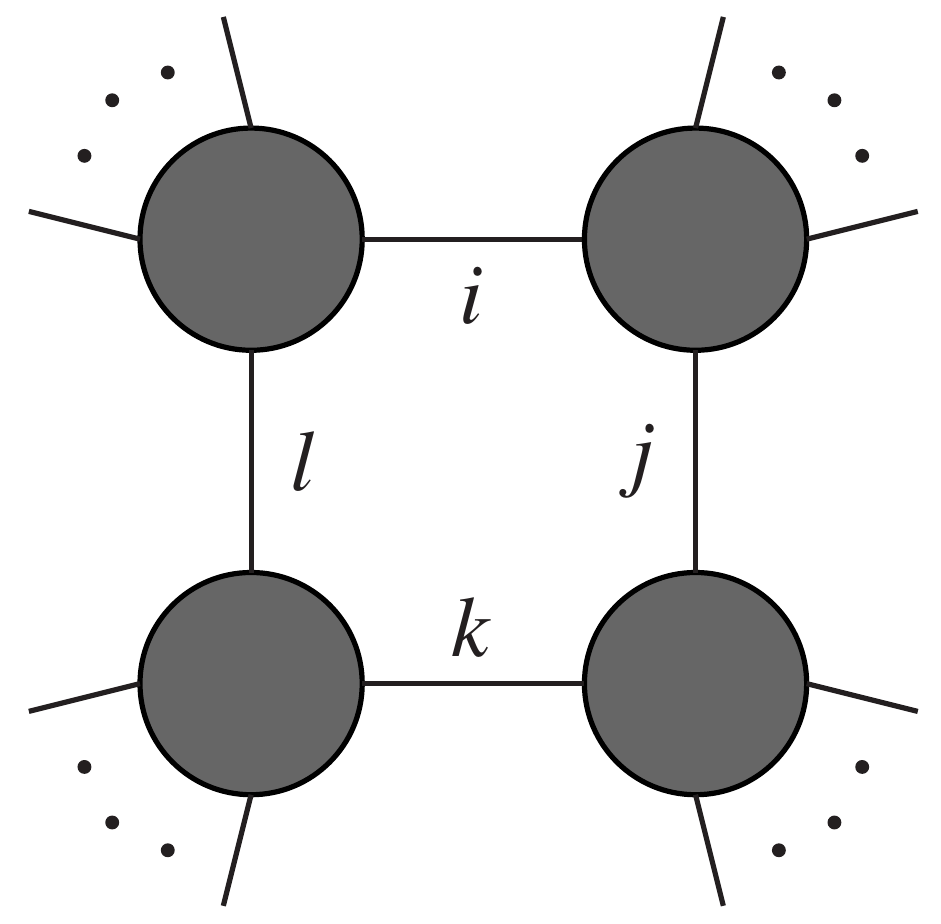}{24}{11}
\end{equation}
%
where the internal loop lines $i,j,k,l$ are on-shell.
The on-shell external lines at each of the blobs, represented by the dots, are uniquely determined by the choice of the internal lines $i,j,k,l$.
The four blobs represent tree-level on-shell amplitudes, each of them with two extra on-shell lines $i,j$ and $j,k$ etc., and we see that, for the determination of the coefficients of $\mathrm{Box}_{ijkl}$, the LHS of \Equation{integranddecom} can be evaluated by sewing together four on-shell tree-level amplitudes.
The same can be done for triangles with three blobs, and for bubbles with two blobs.
In the latter case, the procedure is equivalent to cutting the loop amplitude within the classic application of unitarity, hence the name.

\section{One-loop amplitudes with a space-like gluon}
A one-loop amplitude with an external auxiliary quark anti-quark pair with momenta $p_{A'}^\mu=k^\mu-p_{A}^\mu$ and $p_{A}^\mu$ from \Equation{Eq:pA}, which are diverging linearly with $\Lambda$, has denominators with denominator momentum $K_i^\mu=p_A^\mu+\bar{K}_i^\mu$ containing these momenta.
We restrict ourselves to sectors of the amplitude with at least one such denominator: other contributions do not need any special regarding the diverging momenta.
Each such denominator has a factor $\Lambda$ associated with it in the numerator of the integrand, and one could evaluate the integrand by taking
%
\begin{equation}
\frac{\Lambda}{(\ell+p_A+\bar{K})^2} \;\;\overset{\Lambda\to\infty}{\to}\;\;
\frac{1}{2\lop{p}(\ell+\bar{K})}
\label{denominatorlimit}
\end{equation}
%
which is equivalent with evaluating the integrand using eikonal Feynman rules for the auxiliary quark line.
The integrand has an overall factor $\Lambda$, like the tree-level amplitudes, stemming from the external spinors $\leftA{p_{A'}}\to\sqrt{\Lambda}\,\leftA{p}$ and $\Srght{p_{A}}\to\sqrt{\Lambda}\,\Srght{p}$.

It turns out that one cannot apply the integrand methods on this integrand and evaluate the master integrals by simply reading \Equation{denominatorlimit} backwards.
The reason is the anomalous behavior of bubbles
%
\begin{equation}
\int\frac{\dDell}{(\ell+K_0)^2(\ell+p_A+K_1)^2}
= -\log\Lambda + \frac{1}{\varepsilon} + 2 + \log\left(\frac{2\lop{p}{(K_1-K_0)}}{-\mu^2}\right)
~,
+\Ord(\varepsilon)
\end{equation}
%
and the anomalous triangle
%
\begin{equation}
\int\frac{\dDell}{(\ell^2)(\ell+p_A)^2(\ell+p_A+p_{A'})^2}
=
\frac{1}{k_T^2}\left\{
  \frac{1}{\varepsilon^2}
 -\frac{1}{\varepsilon}\ln\left(\frac{k_T^2}{-\mu^2}\right)
 +\frac{1}{2}\ln^2\left(\frac{k_T^2}{-\mu^2}\right)
\right\}
+\Ord(\varepsilon)
~.
\end{equation}
%
Their behavior is anomalous in the sense that they do not follow na\"ive power counting in $\Lambda$, and swallow a factor of $\Lambda$ in the denominator after integration.
They do not vanish for large $\Lambda$ while they are missing a factor of $\Lambda$ in the numerator compared to \Equation{denominatorlimit}.
This indicates the possible danger of the one-loop amplitude to behave power-like in $\Lambda$, but in~\cite{vanHameren:2017hxx} it is shown that
\begin{enumerate}
\setcounter{enumi}{-1}
\item in the limit of \Equation{limit}, the one-loop amplitudes diverges at most as $\log^2\Lambda$.
\end{enumerate}
It is further explained that the one-loop amplitudes can be calculated as follows:
\begin{enumerate}
\setcounter{enumi}{0}
\item Choose a momentum routing such that the divergent components of the auxiliary quark flow stricktly through the internal auxiliary quark line.
\end{enumerate}
One could shift the integration momentum and let those components also flow through internal gluons, but the above is more practical.
\begin{enumerate}
\setcounter{enumi}{1}
\item Only contributions from master integrals with at most {\em one\/} denominator from the auxiliary quark line need to be considered.
\end{enumerate}
Contributions from master integrals with more such $\Lambda$-dependent denominators vanish, or in case of the box with two of them, decompose into lower-point master integrals with only one such denominator.
Expressions for all master integrals with one such denominator can be found in~\cite{vanHameren:2017hxx}.
\begin{enumerate}
\setcounter{enumi}{2}
\item The contribution from all box and triangle master integrals, except the anomalous triangle, can be obtained by applying integrand methods on the integrand at the limit $\Lambda\to\infty$, which can be constructed directly using eikonal Feynman rules.
\end{enumerate}
So for these boxes and triangles the most straightforward method to calculate the amplitude can readily be applied.
The solutions for the loop momentum $\ell^\mu$ to the cut equations setting the necessary denominators to zero can directly be extrapolated from the usual solutions provided in literature.
\begin{enumerate}
\setcounter{enumi}{3}
\item The residue of the LHS of \Equation{integranddecom} necessary to obtain the contribution from the anomalous triangle is given by an amplitude with an auxiliary gluon pair with divergent momenta.
\end{enumerate}
Instead of assigning divergent momenta $p_A^\mu,p_{A'}^\mu$ to an auxiliary quark-antiquark pair, one could wonder what happens if one assigns them to a pair of gluons.
This is exactly what is needed for the anomalous triangle coefficients, and turns out to be completely equivalent to an amplitude with an auxiliary quark antiquark pair.
Coincidentally, this gives a positive answer to he question whether the original tree-level amplitudes with space-like gluons can be defined with the help of an auxiliary gluon pair.

In general, one may wonder what happens if several external partons carry momenta with divergent components while their sum is finite.
It turns out that it is indeed necessary to study more cases than only the quark-antiquark case in order to tackle the bubble contribution.
\begin{enumerate}
\setcounter{enumi}{4}
\item The residue of the LHS of \Equation{integranddecom} necessary to obtain the bubble contributions are obtained by sewing together tree-level amplitudes with a quark-antiquark pair {\em and\/} a gluon with diverging momentum components.
\end{enumerate}
These tree-level amplitudes can still be calculated with the help of eikonal Feynman rules, but behave, before dividing by $\Lambda$ as dictated by \Equation{limit}, as $\sqrt{\Lambda}$ rather than $\Lambda$, compensating for the anomalous behavior of the bubbles.
The rational terms related to the sector of on-loop graphs with auxiliary quark propagators in the loop, finally, can be obtained using the spurious bubble coefficients.

\section{Summary}
The approach of~\cite{vanHameren:2012if} to the definition and calculation of tree-level amplitudes with a space-like gluon can be generalized to one-loop amplitudes.
The external space-like gluon is interpreted as an external auxiliary quark-antiquark pair.
A parameter, expressing a freedom in the choice of the momenta assigned to the quark-antiquark pair, acts as the regulator of divergencies due to linear denominators.
The method is manifestly gauge invariant and Lorentz covariant, and admits efficient integrand methods for the calculation of one-loop amplitudes.

\providecommand{\href}[2]{#2}\begingroup\raggedright\endgroup

\end{document}